\documentclass{article}

\usepackage{arxiv}

\usepackage[utf8]{inputenc} 
\usepackage[T1]{fontenc}    
\usepackage{hyperref}       
\usepackage{url}            
\usepackage{booktabs}       
\usepackage{amsfonts}       
\usepackage{nicefrac}       
\usepackage{microtype}      
\usepackage{lipsum}
\usepackage{graphicx}
\graphicspath{ {./images/} }

\usepackage{xurl}
\usepackage{xcolor}

\title{Large Language Models in Software Documentation and Modeling: A Literature Review and Findings}

\author{
 Lukas Radosky\textsuperscript{0000-0003-3909-3219} \\
  Department of Applied Informatics\\
  Faculty of Mathematics, Physics and Informatics\\
  Comenius University Bratislava\\
  Bratislava, Slovakia \\
  \texttt{lukas.radosky@fmph.uniba.sk} \\
   \And
 Ivan Polasek\textsuperscript{0000-0001-6004-701X} \\
  Department of Applied Informatics\\
  Faculty of Mathematics, Physics and Informatics\\
  Comenius University Bratislava\\
  Bratislava, Slovakia \\
  \texttt{ivan.polasek@fmph.uniba.sk} \\
}

\begin{document}
\maketitle

\noindent\colorbox{yellow!20}{\parbox{\linewidth}{
  This is a preprint of a paper that was presented at the \textbf{IEEE 24th World Symposium on Applied Machine Intelligence and Informatics (SAMI 2026)}.
}}

\begin{abstract}
Generative artificial intelligence attracts significant attention, especially with the introduction of large language models. Its capabilities are being exploited to solve various software engineering tasks. Thanks to their ability to understand natural language and generate natural language responses, large language models are great for processing various software documentation artifacts. At the same time, large language models excel at understanding structured languages, having the potential for working with software programs and models. We conduct a literature review on the usage of large language models for software engineering tasks related to documentation and modeling. We analyze articles from four major venues in the area, organize them per tasks they solve, and provide an overview of used prompt techniques, metrics, approaches to human-based evaluation, and major datasets.
\end{abstract}


\section{Introduction}

Large language models (LLMs) ignited previously unseen interest in the area of software engineering (SE). Many shortcomings of rule-based, machine learning, and deep learning approaches, such as the need for careful rule or hyperparameter setup and reliance on specific input formats, were instantly overcome. Papers concerned with LLMs for SE (LLM4SE) report that their LLM-based approaches to SE tasks outperform previous state-of-the-art approaches even with simple zero-shot prompts~\cite{24_Tao_KADEL_commit,2024_chen_git_titleCompletion,2024_yang_stackOverflow_AutomaticBiModalQuestionTitleGeneration}.

LLM4SE research entails tasks across the entire software development life cycle (SDLC), including requirements engineering~\cite{2025_SLR_RequirementEngineeringAndSWTesting,2025_SLR_requirements,2025_SLR_requirements_02,2025_voria_RECOVERTowardRequirementsGenerationFromStakeholdersConversations,2025_schwedt_CruiseAC,2025_devito_LLMBasedAutomationOfCOSMICFunctionalSizeMeasurementFromUseCases}, software design and modeling~\cite{2025_SLR_DesignArchitecture,2024_huang_RevealingTheUnseenAIChainOnLMMs4PredictingImplicitDataflows,2024_tinnes_SwModelEvolutionOfLLMs,2025_burgue_AutomationInModelDrivenEngineeringALookBackAndAhead}, implementation~\cite{2025_SLR_CodeGeneration_01,2025_SLR_CodeGeneration_02,2025_SLR_IDE}, and maintenance~\cite{2025_DetectingAndCorrectingSmells}.

The vast amount of LLM4SE research papers also led to many systematic literature reviews (SLRs), with some comprehensively analyzing the entire area~\cite{2024_SLR_TheBest,2024_SLR_Generic_01,2024_SLR_Generic_02,2024_SLR_Generic_03} while others focus on a narrow topic or aspect of LLM4SE~\cite{2024_SLR_Metrics,2025_SLR_NoviceDeveloperPerspective,2025_SLR_productivity}. Our research is focused on tasks related to software documentation and modeling. Relevant SLRs in this area focus on a specific subarea, such as software architecture~\cite{2025_SLR_DesignArchitecture}. We conducted a literature review on software documentation and modeling in a very broad sense, e.g. we include the code summarization task, as the generated summary serves as documentation for the analyzed code.

The remainder of this paper is structured as follows. Section~\ref{sec:RelatedWork} reviews existing SLRs on LLM4SE. Section~\ref{sec:Methodology} describes our approach to discovering relevant literature. Section~\ref{sec:Findings} provides an overview of the analyzed literature across relevant SE tasks. Section~\ref{sec:MetricsandEvaluation} discusses the dominant metrics observed in the analyzed literature and human-based evaluation approaches. Section~\ref{sec:Conclusion} concludes the contributions of this paper and possible research directions.

\section{Related work}
\label{sec:RelatedWork}

Having attracted considerable attention, LLM4SE has been subject to many SLRs. Perhaps the most comprehensive SLR in this area~\cite{2024_SLR_TheBest} analysed tasks, datasets, metrics, architectures, and tuning techniques of LLM4SE across all SDLC phases. Other SLRs concerned with LLM4SE without a more specific focus include~\cite{2024_SLR_Generic_01,2024_SLR_Generic_02,2024_SLR_Generic_03}. Several SLRs considered research across all SDLC phases, focusing on a specific approach or aspect of LLMs for SE, such as multi-agent systems (MASs)~\cite{2024_SLR_AgentiAI4SE,2025_SLR_AgentAI4SE,2025_SLR_AgenticAI4SE_02}, comparison of LLMs and MASs~\cite{2025_SLR_LLMsVsAgenticAI}, benchmarks for code LLMs~\cite{2025_SLR_CodeLLMBenchmarks},  explainability~\cite{2025_SLR_Explainability}, prompt engineering~\cite{2025_SLR_PromptEngineering}, metrics~\cite{2024_SLR_Metrics}, vulnerability detection~\cite{2025_SLR_VulnerabilityDetection}, developers' productivity~\cite{2025_SLR_productivity}, novice developers' perspectives~\cite{2025_SLR_NoviceDeveloperPerspective}, and utilization in SE education~\cite{2024_SLR_SEEducation}.

Some SLRs focus on a specific SDLC phase, including requirements engineering~\cite{2025_SLR_RequirementEngineeringAndSWTesting,2025_SLR_requirements,2025_SLR_requirements_02}, software design with a focus on architectures~\cite{2025_SLR_DesignArchitecture}, implementation and programming~\cite{2025_SLR_CodeGeneration_01,2025_SLR_CodeGeneration_02} with a focus on IDEs~\cite{2025_SLR_IDE} or teaching~\cite{2025_SLR_TeachingProgramming}, and testing~\cite{2025_SLR_RequirementEngineeringAndSWTesting,2025_SLR_Testing_01,2025_SLR_Testing_02}.

\section{Methodology}
\label{sec:Methodology}

In the first phase, we identified four venues for our analysis, IEEE Transactions on Software Engineering (TSE), ACM Transactions on Software Engineering and Methodology (TOSEM), Springer Empirical Software Engineering (EMSE), and International Conference on Software Engineering (ICSE). This ensured the analyzed papers would be related to SE. Due to rapid advances in LLM-related research, we focused on publications from the years 2024 and 2025. All publications from the identified years and venues were manually searched. The second phase involved a brief inspection of the title and abstract as well as a full-text search for keywords to identify LLM-related papers. The keywords involved \textit{LLM}, \textit{language model}, \textit{GPT}, \textit{BERT}, and some variations, such as \textit{large LM} or \textit{large code model}. Searching for the \textit{GPT} keyword as potentially the most popular LLM (and \textit{BERT} as its predecessor) was a heuristic to avoid omitting LLM-related papers that do not explicitly use words such as \textit{LLM}. Finally, the third phase involved manually reading abstracts (and partially full text, when in doubt) to identify papers concerned with software documentation and modeling. We applied a very broad definition, also involving papers concerned with tasks such as code summarization, as the generated summary documents the code in a sense.

\section{Findings}
\label{sec:Findings}

\begin{figure}[htbp]
\includegraphics[width=0.90\textwidth]{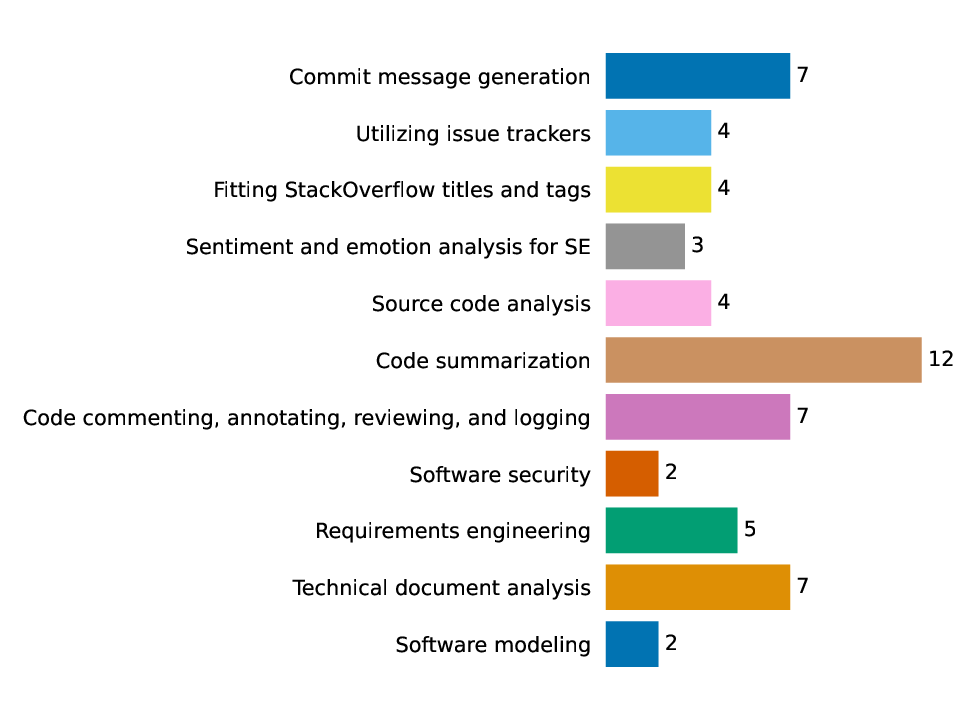}
\caption{Analyzed paper counts per task categories.} \label{fig_category_counts}
\end{figure}

We categorize the analyzed papers based on the tasks they solve. Paper counts per category are shown in Fig.~\ref{fig_category_counts}.

\subsection{Commit message generation}
The commit message generation (CMG) task is concerned with generating textual summarization of differences between two versions of code for purposes of a source code version control system. The research in this area has yielded novel approaches, such as the novel learning method KADEL~\cite{24_Tao_KADEL_commit}, the specialized model CommitBART~\cite{2024_Liu_CommitBART}, or the full-fledged pipeline OMEGA~\cite{25_Imani_ContextConquersParameters_Commits}. Other papers are involved with utilizing existing LLMs and benchmarking various prompting techniques and models instead. Xue et al.~\cite{24_Xue_AutomatedCommitMessageGenrationWithLLMs} performed benchmarking across various LLMs, Wang et al.~\cite{25_Wang_IsItHardToGenerateHolisticCommitMessage} focused on generating holistic, i.e. more detailed commit messages with various models and approaches, Imani et al.~\cite{25_Imani_ContextConquersParameters_Commits} compared open-source LLMs with proprietary models, and Wu et al.~\cite{2025_wu_empiricalstudycommitmessage} comprehensively compared various models and prompting techniques.

Most papers use zero-shot prompting, either per guidelines~\cite{24_Tao_KADEL_commit}, to mitigate prompt influence on model comparison~\cite{25_Imani_ContextConquersParameters_Commits}, to mimic practical usage~\cite{24_Xue_AutomatedCommitMessageGenrationWithLLMs}, or due to the expensive API of the GPT models~\cite{25_Wang_IsItHardToGenerateHolisticCommitMessage}. Few-shot prompting was used as a baseline to own approach~\cite{2024_Liu_CommitBART} or in early model comparison~\cite{25_Imani_ContextConquersParameters_Commits} model comparison. Wu et al.~\cite{2025_wu_empiricalstudycommitmessage} comprehensively compared various models and prompting techniques, hinting that few-shot prompting mitigates the influence of the prompt and that the RAG approach improves, whereas too many examples hinder the model's performance.

The most popular dataset for the CMG task appears to be the MCMD~\cite{2021_tao_MCMD} dataset, used by several papers both in unaltered~\cite{24_Tao_KADEL_commit} and enhanced~\cite{24_Xue_AutomatedCommitMessageGenrationWithLLMs,2025_wu_empiricalstudycommitmessage} forms. Most often, dataset usage is accompanied by the usage of custom collected data.

\subsection{Utilizing issue trackers}

Issue trackers (e.g. GitHub) contain valuable data that is utilized in various ways. Nikoo et al.~\cite{2024_nikoo_git_bmpnModels} explored BPMN models present in open-source GitHub repositories. They used the GPT-4 model with zero-shot prompting to classify BPMN models into domains. LLM-based classification (promptless models) was also utilized by Abedini et al.~\cite{2024_abedini_git_appReviewClassification} to classify issues from issue-tracking systems (such as GitHub) into bug report and feature request categories, to further transform them into app-review-like forms. Yang et al.~\cite{2024_yang_APreliminaryInvestigationOnUsingMultiTaskLearningToPredictChangePerformanceInCodeReviews} proposed a multi-task learning approach for the code change performance prediction task based on code reviews. They evaluated the predictive capabilities of eight (promptless) models. Widyasari et al.~\cite{2025_Widyasari_ExplainingExplanationsAnEmpiricalStudyOfExplanationsInCodeReviews} utilized LLMs with zero-shot prompts to generate an explanation of a specified type for a given Gerrit issue.

\subsection{Fitting StackOverflow titles and tags}

StackOverflow (SO) is a popular platform for discussing software development issues. Its popularity and vastness grant researchers' attention.

Chen et al.~\cite{2024_chen_git_titleCompletion} fine-tuned the CodeT5 model for generating fitting titles for SO posts and GitHub issues and compared it with several other models. The problem of generating titles for SO posts was also studied by Yang et al.~\cite{2024_yang_stackOverflow_AutomaticBiModalQuestionTitleGeneration}. They introduced the SOTitle+ approach combined with a fine-tuned CodeT5 model for title generation. Both papers used zero-shot prompting. He et al.~\cite{2024_he_PTM4TagPlusTagRecommendationOfSOPostsWithPretrainedModels} introduced PTM4Tag+, a framework for recommending tags for SO posts that uses an LLM (CodeT5) as a classifier. Gong and Zhang~\cite{2024_gong_MR2KG} proposed a RoBERTa-based solution for creating knowledge graphs from SO posts, to be used in question answering.

\subsection{Sentiment and emotion analysis for SE}

SO posts, GitHub issues, and other records in SE-related systems may entail emotional charge, i.e. sentiment.
Cassee et al.~\cite{2024_cassee_TranformersAndMetatokenizationInSentimentAnalysisForSE} benchmarked existing sentiment analysis tools. Zhang et al.~\cite{2025_zhang_RevisitngSentimentAnalysisForSEInTheEraOfLLMs} advanced the sentiment analysis for software engineering (SA4SE) task by employing LLMs, distinguishing between general bigger LLMs (bLLMs) and fine-tuned smaller LLMs (sLLMs) and using various prompting techniques (zero-shot and few-shot). Imran et al.~\cite{2024_Imran_UncoveringTheCausesOfEmotionsInSWDeveloperCommunication} used zero-shot LLMs for detecting emotions in software developer communications, which were outperformed by BERT-based models and other baselines.

\subsection{Source code analysis}

Analyzing source code is concerned with information extraction from the given source code snippet. The required information type largely varies.

Several researchers pursued very specific tasks, yielding a chain-of-thought approach for generating dataflow graphs from Python source code~\cite{2024_huang_RevealingTheUnseenAIChainOnLMMs4PredictingImplicitDataflows}, KNN-classified LLM-generated source code embeddings for design pattern recognition~\cite{2025_pandey_DesignPatternRecognitionAStudyOfLLMs}, scientific notebook assessment driven by DistilBERT~\cite{2025_Ghahfarokhi_PredictingTheUnderstandabilityOfComputationalNotebooksThroughCodeMetricsAnalysis}, and the binary functionality classification (BFC) solution PromeTrans~\cite{2024_sha_PromeTransBootstrapBinaryFunctionalityClassification}.

\subsection{Code summarization}

Code summarization is concerned with explaining a snippet of code by concise text. LLMs' generative capabilities are a perfect fit for this task, attracting much attention.

This area is rich with novel approaches and frameworks, including MODE-X~\cite{2024_saberi_UtilizationOfPretrainedLMsForAdapterBasedKnowledgeTransferInSE} model created by introducing an adapter module for enhancing pre-trained models, extractive-and-abstractive EACS framework~\cite{2024_sun_AnExtractiveAndAbstractiveFrameworkForSourceCodeSummarization}, usage of function signatures on input~\cite{2024_ding_DoCodeSummarizationModelsProcessTooMuch}, Esale~\cite{2024_fang_Esale} approach defined by modifying encoder models to capture task-specific code features, StructCodeSum~\cite{2024_zhou_LearningToGenerateStructuredCodeSummariesFromHybridCodeContext} approach focused on producing Javadoc-like summaries, ACE framework~\cite{2025_fang_AHolisticApproach2DesignUnderstandingThroughConceptExplanation}, semantic prompt augmentation~\cite{2024_Toufique_AutomaticSemanticAugmentationOfLanguagePrompts}, HINT approach~\cite{2024_Gao_HINT} an IDE plugin~\cite{2024_Daye_UsingAnLLMToHelpWithCodeUnderstanding}, and novel SIDE metric~\cite{2024_SIDE}.

Zhu et al.~\cite{2024_zhu_DeepIsBetterEmpiricalComparisonOfIRAndDLApproachesToCodeSummarization} performed a benchmark on existing models and approaches for the source code summarization task. Where applicable, they used simple zero-shot prompting. Sun et al.'s benchmark~\cite{2025_sun_SrcCodeSummarizationInEraOfLlms} focused on prompts and compared various prompting techniques (zero-shot, few-shot, chain-of-thought, etc.) and models.

Based on popularity~\cite{2024_zhu_DeepIsBetterEmpiricalComparisonOfIRAndDLApproachesToCodeSummarization,2025_sun_SrcCodeSummarizationInEraOfLlms,2024_zhou_LearningToGenerateStructuredCodeSummariesFromHybridCodeContext,2024_fang_Esale,2024_sun_AnExtractiveAndAbstractiveFrameworkForSourceCodeSummarization,2024_saberi_UtilizationOfPretrainedLMsForAdapterBasedKnowledgeTransferInSE,2024_Toufique_AutomaticSemanticAugmentationOfLanguagePrompts}, the most established datasets for code summarization are CodeSearchNet~\cite{2020_CodeSearchNet,2021_Lu_CodeXGlue}, and the PCSD~\cite{2017_Barone_PCSD} dataset used by~\cite{2024_zhu_DeepIsBetterEmpiricalComparisonOfIRAndDLApproachesToCodeSummarization,2024_sun_AnExtractiveAndAbstractiveFrameworkForSourceCodeSummarization,2024_fang_Esale}.

\subsection{Code commenting, annotating, reviewing, and logging}

Generating comments, annotations, reviews, and logging suggestions from source is similar to code summarization in the need to comprehend code and generate text corresponding to it. However, the difference lies in various purposes and expectations from the generated text.

Among the studied approaches to comment generation are TG-CUP~\cite{2025_chen_TGCUP_CommentUpdating}, a Transformer-based model for comment updating upon source code changes, SCGen~\cite{2023_guo_SnippetCommentGenerationBasedOnCodeContextExpansion} utilizing AST of input source code, few-shot prompting~\cite{2024_Geng_LLMsAreFewShotSummarizers}, and usage of bytecode-derived information~\cite{2025_huang_TowardsImprovingPerformanceOfCommentGenerationModels}.

SpecGen~\cite{2025_ma_specgen} is an LLM-driven mutation-based approach for generating formal specifications of given source code as annotations. Open-source LLMs' ability to perform code review has been investigated in~\cite{2024_yu_FineTuningLlmsToImproveAcc}. An interesting idea of suggesting logging texts based on surrounding source code led to the creation of the LoGenText~\cite{2023_ding_LoGenTextPlus} approach.

\subsection{Software security}

Dealing with software security sometimes entails analysis of documents written in natural language (NL). Research in this area yielded a framework~\cite{2025_han_DoChaseYourTail} for enhancing textual vulnerability descriptions of software with a small user base (i.e. long-tail software) based on few-shot predictions and additional verification against hallucinations. Moreover, LLMs with few-shot prompting were used for the task of generating symbolic models from security protocols~\cite{2025_mao_LLMAidedAutomaticModeling4SecurityProtocolVerification}.

\subsection{Requirements engineering}

Requirements engineering often entails informal conversations and loose mail communication in NL which can be effectively analyzed with LLMs. RECOVER~\cite{2025_voria_RECOVERTowardRequirementsGenerationFromStakeholdersConversations} enables generating formal requirements from informal sources with the human-in-the-loop approach. User stories (USs) are entailed in ReFair~\cite{2024_Ferrara_ReFair} framework for sensitive feature classification from USs and CrUISE-AC~\cite{2025_schwedt_CruiseAC} method for issue investigation and acceptance criteria generation for USs. It uses an LLM ensemble identified as the best from multiple evaluated LLM ensembles. CosMet~\cite{2025_devito_LLMBasedAutomationOfCOSMICFunctionalSizeMeasurementFromUseCases} is an LLM-powered tool for automated COSMIC measurement calculation from a set of use cases, enabling faster software size estimation. Fine-SE is a novel model using the BERT model to extract novel expert features from software artifacts~\cite{2024_Yue_FineSE} for software effort estimation.

\subsection{Technical document analysis}

Technical documents are often well-structured, yet they entail NL and complexity, making their analysis a fitting task for LLMs. Research in this area focused on classifying often overlooked exception handling bugs~\cite{2024_silva_TowardsAutomaticLabelingOfExceptionHandlingBugsACaseStudyOf10YearsBugFixingInApacheHadoop}, gathering more detailed information during the software issue reporting process~\cite{2024_aktas_ImprovingTheQualityOfSoftwareIssueReportDescriptionsInTurkishAnIndustrialCaseStudtyAtSoftech}, improving summarization and question answering related to bug reports~\cite{2025_tamanna_ChatGPTInaccuracyMitigationDuringTechnicalReportUnderstandingAreWeThereYet}, and question answering about library choice~\cite{2024_Tanzil_ChatGPTIncorectnessDetectionInSWReview} with a MAS. Each paper performed evaluation on custom data. Huang et al. created the LLM-based ARJE framework for extracting APIs and their relations from the text~\cite{2023_Huang_APIEntityAndRelationJointExtractionFromTextViaDynamicPromptTunedLanguageModel}, then further improved on it by inferring more API information from the data learned by the LLM during training~\cite{2024_huang_LetsDiscoverMoreAPIRelation}. Tileria et al.~\cite{2024_Tileria_DocFlow} created DocFlow framework for working with Android and Google Play Service documentations across updates.

\subsection{Software modeling}

Very few papers of the analyzed literature dealt with software modeling (some~\cite{2024_huang_RevealingTheUnseenAIChainOnLMMs4PredictingImplicitDataflows} were mentioned in better-fitting sections). Tinnes et al.~\cite{2024_tinnes_SwModelEvolutionOfLLMs} explored the ability of LLMs to predict a missing model change from a batch of model changes. They chose a few-shot RAG-based prompting approach. Burgue\~{n}o et al.~\cite{2025_burgue_AutomationInModelDrivenEngineeringALookBackAndAhead} provided a comprehensive overview of the area of model-driven engineering, including possible AI and LLM usages. Among other propositions, they suggest using LLMs for early evaluation of emerging domain-specific languages or for mitigating data scarcity.

\section{Metrics and Evaluation}
\label{sec:MetricsandEvaluation}

\textbf{Classification}-oriented tasks mostly use the well-known metrics such as precision, accuracy, recall, F1 score, and ROC-AUC. Examples of classification tasks include determining BPMN model domains~\cite{2024_nikoo_git_bmpnModels}, software issue categories~\cite{2024_abedini_git_appReviewClassification}, SO post tags~\cite{2024_he_PTM4TagPlusTagRecommendationOfSOPostsWithPretrainedModels}, design patterns in code~\cite{2025_pandey_DesignPatternRecognitionAStudyOfLLMs}, science notebook understandability levels~\cite{2025_Ghahfarokhi_PredictingTheUnderstandabilityOfComputationalNotebooksThroughCodeMetricsAnalysis}, or  exception handling bugs~\cite{2024_silva_TowardsAutomaticLabelingOfExceptionHandlingBugsACaseStudyOf10YearsBugFixingInApacheHadoop}.

The presence of \textbf{regression} tasks is modest, only examples being code review change performance prediction~\cite{2024_yang_APreliminaryInvestigationOnUsingMultiTaskLearningToPredictChangePerformanceInCodeReviews} and software size estimation from use cases~\cite{2025_devito_LLMBasedAutomationOfCOSMICFunctionalSizeMeasurementFromUseCases}. They used common metrics, such as MAE, MdAE, RMSE, and NRMSE.

The prevalent task type appears to be \textbf{text generation} as it exploits the LLMs' primary advantage - their generative capabilities. The prevalent metrics include BLEU~\cite{2002_Papineni_BLEU}, METEOR~\cite{2005_Banerjee_METEOR}, and ROUGE~\cite{2004_Lin_ROUGE}, used for example in CMG~\cite{24_Tao_KADEL_commit,2024_Liu_CommitBART,25_Imani_ContextConquersParameters_Commits,24_Xue_AutomatedCommitMessageGenrationWithLLMs,25_Wang_IsItHardToGenerateHolisticCommitMessage,2025_wu_empiricalstudycommitmessage}, SO title generation~\cite{2024_chen_git_titleCompletion,2024_yang_stackOverflow_AutomaticBiModalQuestionTitleGeneration}, and code summarization~\cite{2025_sun_SrcCodeSummarizationInEraOfLlms,2024_saberi_UtilizationOfPretrainedLMsForAdapterBasedKnowledgeTransferInSE,2024_sun_AnExtractiveAndAbstractiveFrameworkForSourceCodeSummarization,2024_ding_DoCodeSummarizationModelsProcessTooMuch,2024_zhu_DeepIsBetterEmpiricalComparisonOfIRAndDLApproachesToCodeSummarization,2024_fang_Esale,2024_zhou_LearningToGenerateStructuredCodeSummariesFromHybridCodeContext,2024_Toufique_AutomaticSemanticAugmentationOfLanguagePrompts,2024_Gao_HINT} tasks.

Some papers~\cite{2025_sun_SrcCodeSummarizationInEraOfLlms,2025_han_DoChaseYourTail,2025_devito_LLMBasedAutomationOfCOSMICFunctionalSizeMeasurementFromUseCases} also utilize semantic similarity measures such as BERTScore~\cite{2020_BERTSCORE} (text-text pairs) or SIDE metric~\cite{2024_SIDE} (text-code pairs).

Human-based evaluation is performed either by the authors~\cite{2024_tinnes_SwModelEvolutionOfLLMs,2025_Widyasari_ExplainingExplanationsAnEmpiricalStudyOfExplanationsInCodeReviews}, or by invited participants, who are either experts~\cite{24_Tao_KADEL_commit,25_Wang_IsItHardToGenerateHolisticCommitMessage,2024_fang_Esale,2025_schwedt_CruiseAC,2025_ma_specgen} or students in the field~\cite{2024_chen_git_titleCompletion,2025_sun_SrcCodeSummarizationInEraOfLlms}, or a mixture of both~\cite{2025_wu_empiricalstudycommitmessage,2023_ding_LoGenTextPlus,2024_yu_FineTuningLlmsToImproveAcc,2024_Daye_UsingAnLLMToHelpWithCodeUnderstanding}. Usually, several evaluators are deployed, ranging from low counts such as 2-8~\cite{2024_tinnes_SwModelEvolutionOfLLMs,24_Tao_KADEL_commit,25_Wang_IsItHardToGenerateHolisticCommitMessage,2025_wu_empiricalstudycommitmessage,2024_chen_git_titleCompletion,2025_schwedt_CruiseAC,2025_Widyasari_ExplainingExplanationsAnEmpiricalStudyOfExplanationsInCodeReviews,2025_ma_specgen,2024_yu_FineTuningLlmsToImproveAcc} through medium-sized groups (10-15)~\cite{2025_sun_SrcCodeSummarizationInEraOfLlms,2024_fang_Esale} up to larger groups (32-42)~\cite{2024_Daye_UsingAnLLMToHelpWithCodeUnderstanding,2023_ding_LoGenTextPlus}. Human-based evaluation is usually performed on a subset of the data, complemented by applying an LLM-based evaluator on the entire data. In the studied literature, the GPT-4 model is used. Some~\cite{2025_wu_empiricalstudycommitmessage} chose this model per literature recommendations, others~\cite{2025_sun_SrcCodeSummarizationInEraOfLlms} per their own evaluation.

\section{Conclusion and future work}
\label{sec:Conclusion}

We have analyzed 57 papers from selected venues concerned with using LLMs for tasks related to software documentation and modeling. The popular approaches involve promptless models and zero-shot prompting; less attention is given to more advanced techniques such as few-shot prompting and chain-of-thought. While the emerging area of LLM-driven multi-agent systems has been given little attention, several papers mention them as a possible future research direction.

Novel approaches using LLMs are often used to improve existing tasks, e.g. commit message generation, or to improve the structure of existing resources, e.g. StackOverflow tag and title generation. Most papers report improvement over state-of-the-art approaches, often significant. However, the long-awaited revolution stemming from the introduction of LLMs might be lagging behind its hype. So far, the LLMs are being used to improve the speed and quality of existing workflow and task structure in software engineering instead of redefining them. At this point in time, their incorporation seems to lead to evolution rather than revolution. It is yet to be seen whether novel LLM tuning or ensemble techniques will change this trend.

\section*{Acknowledgment}

This research and paper was 100\% funded by the EU NextGenerationEU through the Recovery and Resilience Plan for Slovakia under the project "InnovAIte Slovakia, Illuminating Pathways for AI-Driven Breakthroughs" No. 09I02-03-V01-00029.

\bibliographystyle{unsrt}  
\bibliography{references}

\end{document}